\documentclass[aps,pre,reprint,superscriptaddress]{revtex4-2}

\usepackage[utf8]{inputenc}
\usepackage[T1]{fontenc}
\usepackage{mathptmx}
\usepackage{hyperref}
\usepackage{graphicx,color}
\usepackage{dcolumn}
\usepackage{bm}
\usepackage{amsmath,amssymb}
\usepackage{comment}
\usepackage[bitstream-charter]{mathdesign}
\usepackage{cases}
\usepackage{ulem}
\DeclareSymbolFont{usualmathcal}{OMS}{cmsy}{m}{n}
\DeclareSymbolFontAlphabet{\mathcal}{usualmathcal}

\definecolor{M_Green}{rgb}{0.2 , 0.6 , 0.2}

\newcommand{\be}{\begin{equation}}
\newcommand{\ee}{\end{equation}}
\newcommand{\bea}{\begin{eqnarray}}
\newcommand{\eea}{\end{eqnarray}}

\usepackage{enumitem}

\renewcommand{\leq}{\leqslant}
\renewcommand{\geq}{\geqslant} 

\def\I{\mathrm{i}}                  
\def\D{\mathrm{d}}                  

\def\free{\mathrm{free}}
\def\fixed{\mathrm{fixed}}

\def\eqlaw{ \overset{\mbox{\tiny (law)}}{=}  }

\def\lambdaMinTyp{ \lambda_{1}^\mathrm{typ} }
\begin{document}


\title{Anomalous Scaling of Heterogeneous Elastic Lines: 
\\ A New Picture from Sample to Sample Fluctuations}

\author{Maximilien Bernard}
\email{maximilien.bernard@phys.ens.fr}
\affiliation{LPTMS, Université Paris Saclay, CNRS, 91405 Orsay, France}
\author{Pierre Le Doussal}
\email{ledou@lpt.ens.fr}
\affiliation{Laboratoire de Physique de l’Ecole Normale Supérieure, CNRS, ENS and PSL Université, Sorbonne Université, Université Paris Cité, 24 rue Lhomond, 75005 Paris, France}
\author{Alberto Rosso}
\email{alberto.rosso@universite-paris-saclay.fr}
\affiliation{LPTMS, Université Paris Saclay, CNRS, 91405 Orsay, France}
\author{Christophe Texier}
\email{christophe.texier@universite-paris-saclay.fr}
\affiliation{LPTMS, Université Paris Saclay, CNRS, 91405 Orsay, France}

\date{\today}

\begin{abstract}
We study a discrete model of an heterogeneous elastic line with internal disorder, submitted to thermal fluctuations. The monomers are connected through random springs with independent and identically distributed elastic constants drawn from $p(k)\sim k^{\mu-1}$ for $k\to0$. When $\mu>1$, the scaling of the standard Edwards-Wilkinson model is recovered. When $\mu<1$, the elastic line exhibits an anomalous scaling of the type observed in many growth models and experiments. Here we derive and use the exact probability distribution of the line shape at equilibrium, as well as the spectral properties of the matrix containing the random couplings, to fully characterize the sample to sample fluctuations. Our results lead to novel scaling predictions that partially disagree with previous works, but which are corroborated by numerical simulations. We also provide a novel interpretation of the anomalous scaling in terms of the abrupt jumps in the line's shape that dominate the average value of the observable.
\end{abstract}

\maketitle

\section{Introduction}
\label{sec:intro}


The Edwards-Wilkinson equation \cite{edwards1982surface}, i.e. 
\begin{equation}
\partial_t h(x,t)= k \nabla^2 h(x,t) + \eta(x,t)   
\:,
\end{equation}
is the simplest continuum model for the growth of an interface, of height $h(x,t)$, in presence of
space-time white noise $\eta(x,t)$. By contrast to
other models of growth, such as the Kardar-Parisi-Zhang (KPZ) equation \cite{kardar1986dynamic}, it describes equilibrium growth and satisfies
detailed balance. 
Being linear it is easy to solve, and leads to standard dynamical scaling. This means that starting from
a flat initial condition the interface becomes rougher with a growing length scale $\ell(t) \sim t^{1/z}$
below which is it locally equilibrated, leading to~\cite{family1985scaling} 
\begin{align}
\label{eq:normalscalingx}
 &\text{standard scaling:} 
 \nonumber
 \\
 &
 \langle (h(x,t) - h(0,t))^2 \rangle \sim \begin{cases} \ell(t)^{2 \zeta} =t^{ 2 \beta} \quad ,\quad  \ell(t) < x \\
 x^{2 \zeta} ~~~~~~~~~~~~~~ \quad ,\quad  x < \ell(t) 
 \end{cases} 
\end{align}
We will focus on one space dimension $d=1$, where the 
growth exponent is $\beta = \beta_{\rm EW}= 1/4$, the roughness exponent is $\zeta=\zeta_{\rm EW}= 1/2$ and the dynamical exponent
is $z=z_{\rm EW} = \zeta/\beta=2$. There is another interpretation of this model as describing the 
thermal fluctuations of an elastic line, e.g. its Rouse dynamics \cite{de1976dynamics}, and the EW model is useful in that
context. Here
we will adopt indifferently both interpretations and refer to $h(x,t)$ either as a height field
or as the displacement field of an elastic line, e.g. in a discrete setting,
the position of monomers of a Gaussian chain.

The scaling \eqref{eq:normalscalingx} is quite standard in growth models and holds also for KPZ interfaces. However, in the context of discrete growth models for molecular beam epitaxy (MBE) 
\cite{wolf1990growth,sarma1991new,krug1994turbulent,schroeder1993scaling,sarma1996scale}
a different scaling has been observed: First, the scaling exponents 
are different from EW, although the relations $z = \zeta/\beta$ and $z= 2 \zeta + d$ 
are still valid. \footnote{The latter relation is valid in general space dimension $d$,
and arises since in these MBE models there is an underlying conservation law for the height field.}
More importantly, it was observed that the scaling is {\it anomalous}, i.e.
that it takes the form
\begin{align}
    \label{eq:anomscaltyp}
&\text{anomalous scaling:}  
\nonumber
 \\
 &
 \langle (h(x,t) - h(0,t))^2 \rangle \sim \begin{cases} \ell(t)^{2 \zeta} =t^{ 2 \beta} ~~~~~~   \quad ,\quad  \ell(t) < x \\
 x^{2 \zeta_{\rm loc}} \ell(t)^{ 2 (\zeta - \zeta_{\rm loc})}  \quad ,\quad  x < \ell(t) 
 \end{cases} 
\end{align}
with $\ell(t) \sim t^{1/z}$. There is thus apparently an additional roughness exponent, 
$\zeta_{\rm loc}$, usually called "local" as it appears only in local observables. 
When $\zeta_{\rm loc}=\zeta$ one recovers the standard scaling given in Eq. \eqref{eq:normalscalingx},
as is the case for the EW equation. Another important property which was 
observed is multiscaling behavior, i.e.
\be 
\label{eq:anomscaltypq}
 \langle |h(x,t) - h(0,t)|^q \rangle^{1/q} \sim \begin{cases} \ell(t)^{\zeta} =t^{ \beta} \hspace{1.35cm} ,\quad  \ell(t) < x \\
 x^{\zeta_{\rm loc}(q)} \ell(t)^{ (\zeta - \zeta_{\rm loc}(q))}  ,\quad  x < \ell(t) 
 \end{cases} 
\ee 
These two features have also been observed in growth models \cite{castro1998anomalous} and in a number of experimental systems such as film deposition \cite{auger2006intrinsic,YimThinFilms},
studies involving fracture mechanics in paper sheets
\cite{balakin2000intrinsic,balankin2006intrinsically}
or in granite blocks \cite{lopez1998anomalous}.
In the context of interfaces in 
invasive percolation, it has been shown in Refs.\cite{asikainen2002dynamical,asikainen2002interface} that the
distribution of the height differences display power-law behaviour, and consequently, anomalous and multiscaling behaviour. 
These behaviors have also been observed in imbibition, i.e. fluid flows with quenched disorder 
\cite{soriano2002anomalous, soriano2003anomalous,soriano2005anomalous}, as well as 
in models of these systems \cite{pradas2006intrinsic,pradas2008dynamical}.

A feature common to all these systems is the presence of ripped interfaces, whose instantaneous configurations
exhibit rare large local jumps. This feature, which leads to multiscaling, was compared to intermittency of the velocity field in turbulence \cite{krug1994turbulent}. It thus seems that the existence of a local roughness exponent is characteristic of interfaces dominated 
by large jumps. 

In an important attempt to clarify this anomalous scaling behavior, Lopez et al. 
introduced a variant of the EW equation in one dimension with quenched columnar disorder \cite{lopez1995growth,lopez1996lack,lopez1997superroughening,lopez1999scaling,lopez1997power}
\be 
\label{eq:rdcont}
\partial_th(x,t)=\partial_x\big[k(x)\partial_xh(x,t)\big]+\eta(x,t)
\ee 
where $k(x)$ is a short range correlated random function with a given distribution, and the EW model is recovered for $k(x)=1$.
It is indeed quite natural to ask what is the effect of quenched "internal" disorder on the EW equation. 
Eq. \eqref{eq:rdcont} models an interface where the diffusivity 
or redistribution of growth is inhomogeneous. A new behavior is expected when the diffusion is anomalous. In the elastic line interpretation, the internal disorder means that the elastic matrix
is inhomogeneous and random \footnote{There is a third interpretation in terms
of a population diffusing in a quenched random environment, 
and in presence of random immigration.}. 
An example of heterogeneous elastic manifolds was studied in 
Refs.~\cite{cule1998static,cule1998polymer} 
with the additional presence of a pinning substrate.

In \cite{lopez1995growth,lopez1996lack,lopez1997superroughening,lopez1999scaling,lopez1997power} 
it was shown that if the probability to have a small $k(x)$ is large enough
anomalous scaling arises. More precisely, in these works the authors computed the {\it disorder averaged} (indicated by overline)
height-to-height correlation function,  
$\overline{ \langle (h(x,t) - h(0,t))^2 \rangle }$, and recovered Eq. \eqref{eq:anomscaltyp}. 
Based on the above result they made the claim that this interface
is characterized by two distinct roughness exponents: one that captures the
global shape of the fluctuations of the interface, $\zeta$, and
the other one, $\zeta_{\rm loc}$ which characterizes the local fluctuations. 

In this paper we will reexamine the model \eqref{eq:rdcont} (in a discrete 
version) and show that $\zeta_{\rm loc}$ has in fact a quite different interpretation. As it is often the case for systems with
quenched randomness, the disorder average of some observables may be very different from the typical one for a given piece of interface, in a given disorder realization. Indeed from our results, we find that in each realization a piece of the interface
of size $\ell(t)$ displays a jump of size $\sim t^\beta$.
Hence for $x \ll \ell(t)$ the height difference
$h(x,t) - h(0,t)$ is either typical and negligible (i.e. $\sim x^{ \zeta}$) 
or, with a probability $x/\ell(t)$, contains the abrupt jump of size $\sim t^\beta$. 
This leads to the disorder average \footnote{the translational average is assumed to
be equivalent to a disorder average.}
\be 
\overline{ \langle |h(x,t) - h(0,t)|^q \rangle } \sim \frac{x}{\ell(t)}  t^{q \beta} \quad \text{for} \quad x \ll \ell(t)
\ee 
This formula is valid for  $q>q_c$ where $q_c=1/\zeta<2$ (see below) ; for 
$q=2$ it leads to Eq.~\eqref{eq:anomscaltyp}. In addition it predicts \eqref{eq:anomscaltypq} for $q>q_c$
with 
\be 
\zeta_{\rm loc}(q)  = \frac{1}{q} 
\ee 
For $q< q_c$, the contribution of the large jumps to small moments becomes irrelevant and the standard scaling $\sim x^{q \zeta}$ is recovered.
As a result, the physical interpretation is quite different from Ref. \cite{lopez1995growth,lopez1996lack,lopez1997superroughening,lopez1999scaling,lopez1997power}
since we find that $\zeta_{\rm loc}$ is not a genuine roughness exponent. Instead
the anomalous scaling for this model results from an intermittency phenomenon, quite analogous to
shocks in Burger's turbulence \cite{bernard2000turbulence}. This picture emerges from the
study of the distribution of the observables before disorder averaging and is confirmed by our numerical simulations.
Although our results are derived for the model of equation \eqref{eq:rdcont} 
our formula for $\zeta_{\rm loc}(q)$ appears to agree with the experimental
observations in imbibition \cite{soriano2002anomalous, soriano2003anomalous,soriano2005anomalous,pradas2006intrinsic,pradas2008dynamical} and thus provide a potential 
alternative explanation for the ubiquitous anomalous scaling. 

Our paper is organized as follows. In Section~\ref{sec:model}, we define the discrete version of the model and summarize our main results. We discuss the scaling behaviour for finite interfaces which was overlooked in the litterature. In Section~\ref{sec:boundary} we introduce and implement the different boundary conditions.
In Section~\ref{sec1} we show that, even if the disorder averages of the mean squared displacement, 
$\overline{D(L,t)}$, and height-to-height correlation function,
$\overline{G(x,t)}$ (see definitions below), can grow unboundely with time, the equilibrium distributions of their asymptotic values  $G(x) \equiv G(x,t=\infty)$ and $ D(L) \equiv D(L,t=\infty)$ 
always exist and, for all $\mu$, exhibit power law tails.  In Section~\ref{sec:FTR}
 we provide exact results at finite time based on the average spectral density for the operator $\partial_x k(x) \partial_x$ involved in Eq. \eqref{eq:rdcont}. This model was extensively studied in \cite{lopez1996lack,lopez1997superroughening,lopez1999scaling,lopez1997power,lopez1998anomalous} with different methods and partially different results. In Section~\ref{sec:lopez}, we compare our results to previous works~\cite{lopez1996lack,lopez1997superroughening,lopez1999scaling,lopez1997power,lopez1998anomalous}.
Finally Appendices A-C contain details of the derivation, while
 Appendix D displays the exact result for the equilibrium shape of the interface for a continuum model.

\section{Model and observables} \label{sec:model} 

 We study a modified, inhomogeneous version of the  discrete Edwards Wilkinson (EW) interface  where the  $L+1$ monomers are
connected to their nearest neighbors by  springs
with different constants $k_i>0$.
The interface is initially flat, $h_x(t=0)=0$, and its dynamics is described by  the following set of coupled Langevin equations
\begin{align} 
\label{eqmotion}
\partial_t h_x(t) 
= 
&- k_{x} [h_{x}(t) - h_{x-1}(t)] 
\\\nonumber
&- k_{x+1} [h_{x}(t) - h_{x+1}(t)] 
+ \eta_x(t) 
\end{align}
Here $\eta_i(t)$ are independent Gaussian white noises, $\langle \eta_i(t)\rangle=0$,  
$\langle \eta_i(t)\eta_j(t')\rangle= 2 T \delta_{i,j}\delta(t-t')$.
The brackets $\langle \ldots \rangle$ denote the average over the thermal noise, $T$ being the temperature that we set to unity. 
This is the discrete version of the model of Eq. \eqref{eq:rdcont}.

\begin{figure}[!ht]
\centering
\includegraphics[width=0.5\textwidth]{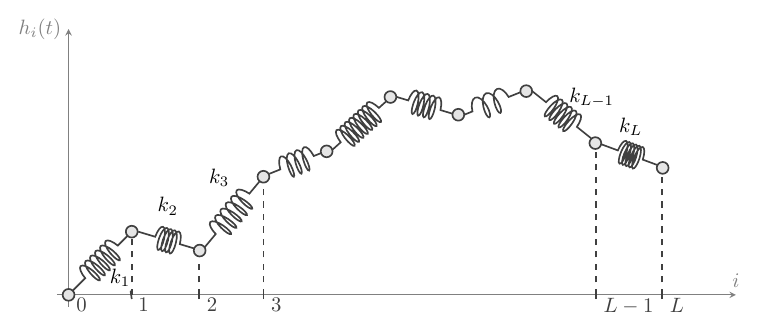}
    \caption{Illustration of a random spring chain for $L$ monomers fixed at the origin and free on the other hand.
    }
    \label{schema}
\end{figure}

We are here interested in characterizing the geometrical properties of the interface:
for this purpose we introduce two quantities
\begin{itemize}
    \item the mean  square displacement  $D(L,t) = \langle \sum_i  h_i^2(t) \rangle /L$ which probes the fluctuations of the whole interface.  
    \item  the height-to-height correlation function $G_i(x,t) = \langle (h_{i+x}(t) - h_i(t))^2 \rangle$, which measures the fluctuations over a distance $x$.
\end{itemize}
 
Working on a finite interval $x\in[0,L]$, we have to specify some boundary conditions at both ends of the line. 
For reasons explained below, we consider that the line is attached at the origin, which corresponds to a Dirichlet boundary condition for the field, $h_0(t)=0$.
The other end of the line is either free (Fig.~\ref{schema}) or is also fixed (the two cases are denoted below the "free" case of the  "fixed" case, respectively).

\subsection{Aim of the paper and main results}

The main question addressed in the paper is to understand the role of disorder in the elastic line, encoded in the spring constants $k_i$ which are quenched random variables
independent and identically distributed with the probability density function (PDF)
\begin{equation}
p(k) = \mu\, k^{\mu-1} \quad \text{with}\;  \; k \in (0,1)
\label{distspring}
\end{equation}
This model was introduced in Refs.~\cite{lopez1996lack,lopez1997superroughening,lopez1999scaling,lopez1997power}.
 Both quantities $D(L,t)$ and $G_i(x,t)$ at a given distance and a given time are random, as they depend on the particular realization of the spring constants. 
 The physics of the interface crucially depends on the positive parameter  $\mu>0$ that characterizes the behaviour of the distribution for small $k$. It is natural to first consider averages over the disorder, indicated with an overline, $\overline{G_i(x,t)}$ and $\overline{D(L,t)}$. For simplicity we consider the bulk of the interface, far from the boundaries, where the dependence on $i$ of $\overline{G_i(x,t)}$ can be discarded.
  We will see that, depending on $\mu$, two regimes should be distinguished.
 
 \textbf{For} {\boldmath $\mu>1$}, the disorder does not affect the Edwards-Wilkinson scaling. Furthermore, the scaling is independent of the boundary conditions. As in the non disordered case, the dynamics is governed by 
 a growing  length, $\ell(t) \sim t^{1/z_\mathrm{EW}}$ (with $z_\mathrm{EW}=2$), as far as $\ell(t)\ll L$. For distances smaller than  $\ell(t)$, the roughness of the interface grows with   characteristic roughness exponent,  $\zeta_\mathrm{EW}=1/2$,
while it saturates for larger distances. As a consequence, we obtain the scaling:
\begin{align}
    \label{w2}
    &\overline{D(L,t)} \sim
    \begin{cases}
    \quad t^{ 2\beta_\mathrm{EW} } \quad & \text{for } \;  t \ll L^{ z_\mathrm{EW} } 
    \\[0.25cm]
    \quad L^{ 2 \zeta_\mathrm{EW} } \quad & \text{for } \;  t \gg L^{ z_\mathrm{EW} } 
    \end{cases}
\\[0.5cm]
    \label{g2}
    &\overline{ G(x,t)  } \sim
    \begin{cases}
    \quad t^{2 \beta_\mathrm{EW} } \quad & \text{for } \;  t \ll x^{ z_\mathrm{EW} }  
    \\[0.25cm]
    \quad x^{2 \zeta_\mathrm{EW} } \quad & \text{for } \;t\gg  x^{ z_\mathrm{EW} } 
    \end{cases}
\end{align}
where we have assumed $x \ll L$. These scalings can be recovered by posing $\overline{ G(x,t)}$ and   $\overline{D(x,L)  } $ equal to $\ell(t)^{2 \zeta}$, until the length $x$ or the length  $L$ are reached, and the exponent  $\beta_\mathrm{EW} =\zeta_\mathrm{EW}/z_\mathrm{EW} =1/4$.

\begin{widetext}
\onecolumngrid

\begin{figure}[!ht]
\centering
\includegraphics[width=1\textwidth]{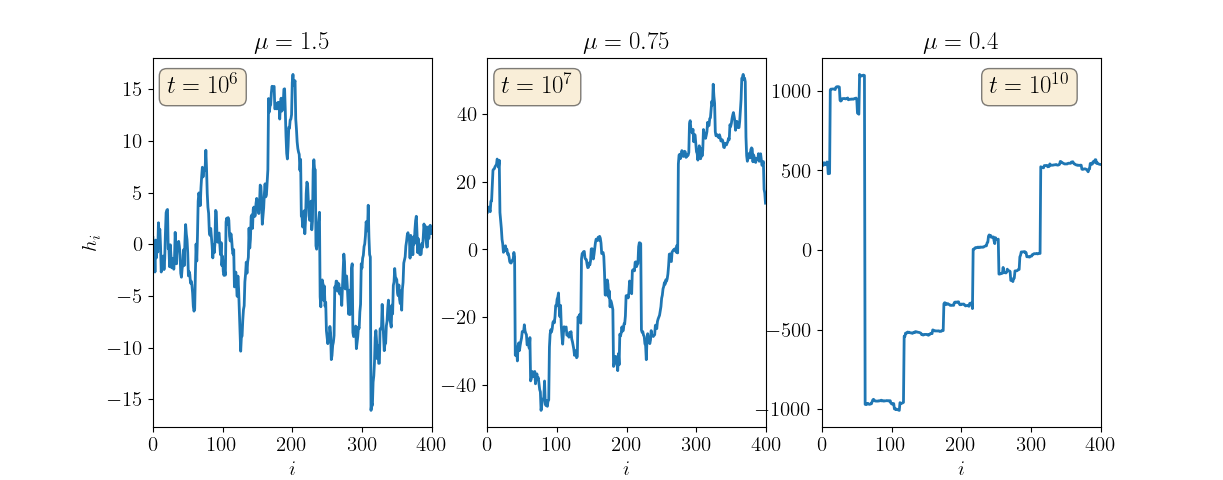}
\caption{Example of three interfaces for $\mu=1.5,  0.75,  0.4$, Dirichlet boundary conditions and zero mean. Decreasing $\mu$, the interfaces display larger and larger jumps which are at the origin of the anomalous behaviour.}
    \label{fig:figint}
\end{figure}
\end{widetext}
\twocolumngrid

\begin{figure}[!ht]
\includegraphics[width=0.5\textwidth]{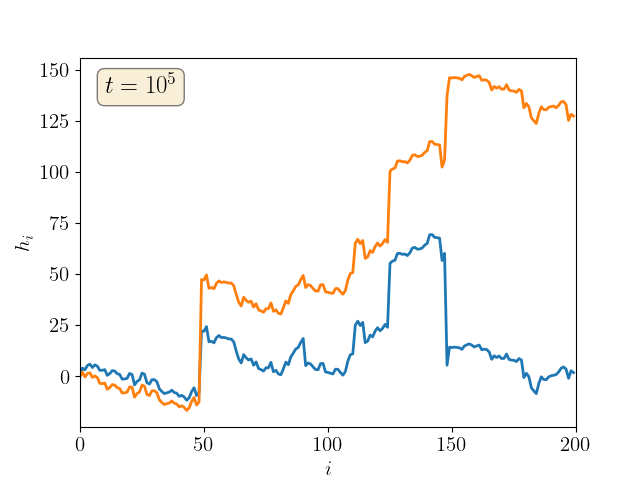}
\includegraphics[width=0.5\textwidth]{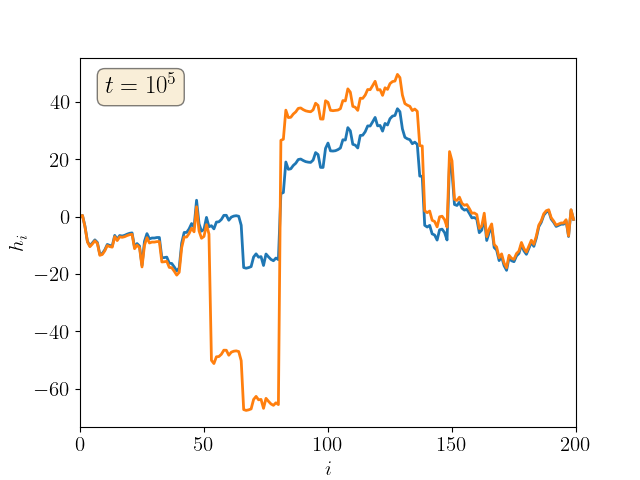}
    \caption{Interfaces ($\mu=0.75$) with a given realization of spring are evolved for a long time $t \sim L^z$ using the same noise. The disorder is typical but the smallest spring value is divided by a factor $10$. Top: fixed case (two Dirichlet boundary conditions) (blue) versus free case (orange). 
    A single anomalously weak spring is enough to rip the interface with free end, but not with two fixed ends. 
    Bottom: fixed case with a single weak spring (blue) versus with two very weak springs (orange, in this case the two smallest springs are divided by a factor $10$). Here we see that only the orange interface is ripped. }
    \label{figintv2}
\end{figure}

\textbf{For} {\boldmath $\mu<1$}, the probability for very small spring constants becomes important and $\overline{1/k_i}$ diverges. In this case we observe  ripped interfaces as shown in Fig. \ref{fig:figint}. 
These sharp jumps are at the origin of a different scaling: 
\\
 (i) The interface is rougher and the dynamics is faster with critical exponents 
 \begin{equation}
  \zeta=1/(2 \mu), \quad   z=(1+\mu)/\mu \quad \text{and} \quad   \beta = 1/[2(1+\mu)].
 \end{equation}
 (ii) the disorder-averaged observables exhibit an anomalous scaling and depend on the boundary conditions.
 \\
\textit{In the free case (one end fixed and the other free)}, we find
\begin{align}
    \label{scaleDmu05free}
  &  \overline{D^\free(L,t)} \sim
    \begin{cases}
    \quad t^{2\beta} \quad & \text{for } \;  t \ll L^z 
    \\[0.25cm]
    \quad  L^\mu t^{1-\mu}  \quad & \text{for } \;  t \gg L^z 
    \end{cases}
    \\[0.5cm]
\label{eq:anomscaling}
&\overline{G^\free(x,t)} \sim
    \begin{cases}
 \quad t^{2 \beta} \quad & \text{for } \;  t \ll x^z   
    \\[0.25cm]
 \quad x t^{(2\zeta-1)/z}  \quad & \text{for } \; x^z \ll t \ll L^z   
    \\[0.25cm]
 \quad x  L^{-1+\mu} t^{1-\mu} \quad & \text{for } \; t \gg L^z
\end{cases}
\end{align}
\textit{In the fixed case} (both ends fixed), we obtain different results at large times $t \gg L^z$: 
\\ 
When $1/2 < \mu <1$, both observable saturate to 
\begin{equation}
  \overline{D^\fixed(L,t)} \sim L^{2 \zeta}
    \hspace{0.5cm}\mbox{and}\hspace{0.5cm}
    \overline{G^\fixed(x,t)} \sim x L^{2 \zeta-1}
\end{equation} 
for $t\to\infty$.
When $ \mu <1/2$, instead, they both diverge as~: 
\begin{align}
\overline{D^\fixed(L,t)} \sim L^{1+2 \mu} t^{1-2 \mu} 
\label{scaleDmu05per} \\[0.25cm]
\overline{G^\fixed(x,t)} \sim x L^{2\mu} t^{1-2\mu}
\label{scalemu05per} 
\end{align}
In this paper we show that this anomalous scaling is due to the fact that the average value of the observables are dominated by rare events. In particular, when $x^z \ll t \ll L^z$, the average $\overline{G(x,t)}$ is dominated by the largest jump within windows of size $\ell(t)$.
The probability that such an event occurs is proportional to the size of the window, i.e. to $x$.
At very large times, the average over disorder is instead dominated by rare interfaces that are ripped by very weak links. Consequences depend on the boundary conditions.
\\
\textit{In the free case}, a single very weak link is enough to rip a piece of the interface (See Fig.~\ref{figintv2}). Hence, the occurrence of such events is controlled by the 
PDF of the weakest link of the chain, which can be easily computed from $p(k)$ (see e.g. \cite{schehr2014exact}):
\begin{equation}
 \label{weakest}
 f_{1,L}(k) = \mu L k^{-1+\mu} e^{-L k^\mu}
    \:.  
\end{equation}
For $\mu<1$, their density diverges 
 close to the origin. As a result,  the mean value of both  $D(L,t)$ and $G(x,t)$ keeps growing even when $t \gg L^z$. \\
\textit{In the fixed case}, at least two very weak link are needed to rip the interface (See Fig.~\ref{figintv2}). Hence,  the occurrence of such rare events is controlled by the PDF of the second weakest link of the chain which is also known from
standard order statistics (see e.g. \cite{schehr2014exact}):
\be
f_{2,L}(k) = \frac{\mu}{2} L^2 k^{-1+2\mu} e^{-L k^\mu}.
\ee
Here, for $\mu<1/2$, their density diverges which implies an unbounded growth of $\overline{G^\fixed(x,t)}$ and $\overline{D^\fixed(L,t)}$ for $t \gg L^z$.


\section{Boundary conditions}
\label{sec:boundary}

The main equation \eqref{eqmotion} can be rewritten as
\begin{equation}
    \label{eqmotion2}
\partial_t h_i(t) = 
- \sum_{j=1}^{L} \Lambda_{i,j}\, h_j(t) + \eta_i(t) 
\hspace{1cm}   
   \text{for} \; \;i=1, \dots , L 
\end{equation}
where $\Lambda$ is a real $L \times L$ symmetric tridiagonal matrix~:

\begin{align}
  & \Lambda_{i,i}=k_{i} + k_{i+1}
   \hspace{2cm}   
   \text{for} \; \;i=1, \dots , L \nonumber \\
  & \Lambda_{i,i+1}=\Lambda_{i+1,i}= -k_{i+1} 
  \hspace{.75cm} 
  \text{for} \;\; i=1, \dots , L-1 \nonumber
\end{align}
At the edge of the interface, it is natural to consider  two types of boundary conditions: the interface is either fixed (e.g. $h_0=0$)  or free, which leads to three cases  (fixed/fixed, fixed/free or free/free). Additionally, one could also consider a fourth case for periodic boundary conditions ($h_0 \equiv h_L$).

 To control the fluctuations of the center of mass, we
 have found convenient to fix the position of one extremity, $h_0=0$, setting there  a Dirichlet boundary condition (i.e. setting $k_1>0$ in the matrix $\Lambda$). 
 For the second extremity we consider two cases: 
 \begin{enumerate}[label={(\arabic*)},leftmargin=*,align=left,itemsep=0.1cm]
     \item 
       The "free case", in which the extremity $h_L$ is free (setting $k_{L+1}=0$  in the matrix $\Lambda$), corresponding to a Neumann boundary condition. 
     \item 
     The "fixed case", in which the second end is also fixed at the origin,  corresponding to a second Dirichlet boundary condition $h_{L+1}=0$ (setting $k_{L+1}>0$). 
 \end{enumerate}
As a result, in both cases, the matrix $\Lambda$ is invertible.

Here, we are mostly interested in scaling properties of the elastic lines, hence, boundary conditions are not expected to lead to dominant effects. Nonetheless, in the presence of anomalous roughening, we have identified two types of behaviours depending on boundary conditions, which are both covered by the two situations considered here. 
Furthermore, our analysis is largely based on analytical expressions for the two-point correlation function $\langle h_i h_j \rangle$ obtained in the two situations.

\begin{figure}[!ht]
\centering
\includegraphics[width=0.45\textwidth]{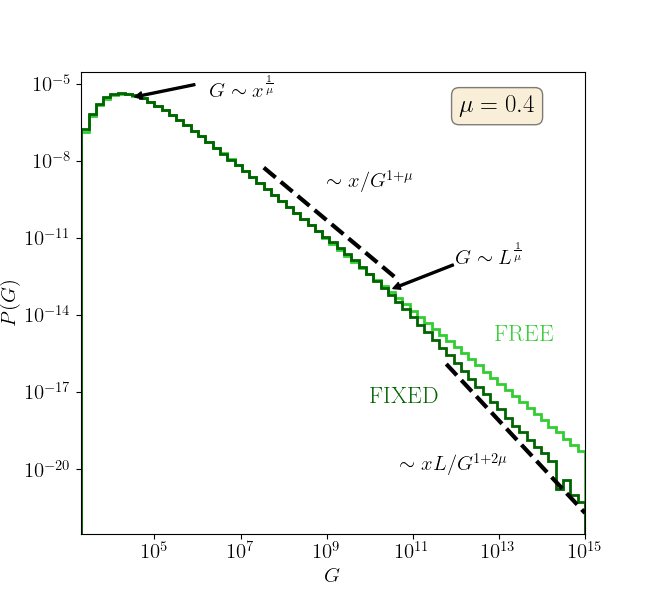}
\includegraphics[width=0.45\textwidth]{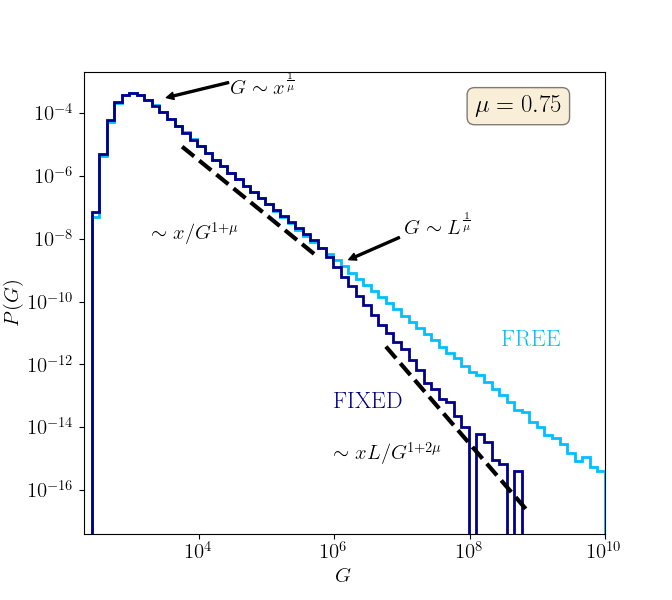}
    \caption{
    Histogram of $G(x)$ for $x\ll L$ in the equilibrated regime for $\mu=0.4$  (Top) and $\mu=0.75$ (Bottom). $N_s=10^7$ values of $G$ are  obtained by direct evaluation of  Eq.~\eqref{Gavg} and Eq.~\eqref{eq:Gfixed}, with $L=10^4$ and $x=100$, in the free case (light green, light blue) and the fixed case (dark green, dark blue). 
    For each distribution, the typical behaviour is located around $G \sim x^{1/\mu}$ ($2 \zeta=1/\mu$). The slow power law decay $\sim x/ G ^{1+\mu}$ as well as the fast  power law decay, for the fixed case, $\sim xL/ G ^{1+ 2\mu}$ are well visible.
     In this latter case, the crossover occurs for $ G  \sim L^{1/\mu}$.} 
    \label{fig:distgasympt}
\end{figure}


\section{Equilibrium distributions: Inverting \texorpdfstring{$\Lambda$}{TEXT}}
\label{sec1}

For the process \eqref{eqmotion},
and for a given realization of springs, the thermal fluctuations are Gaussian and the explicit expression of the equal time two-points correlation function is known  \cite{gupta2013dynamics}:
\begin{equation}
\label{twopointsSDt}
\langle h_i(t) h_j(t) \rangle = \left(\frac{1 - e^{-2 \Lambda t}}{\Lambda}\right)_{ij}.
\end{equation}
In the long time limit, since $\Lambda$ has a strictly positive spectrum, the correlator becomes time independent. 
The equilibrium correlations are given by:
\begin{equation}
\label{twopointnSDt}
\langle h_i h_j \rangle = ({\Lambda}^{-1})_{ij}, \quad \text{for }\, t\to \infty.
\end{equation}
The inversion of the tridiagonal matrix $\Lambda$ can be done using a recursion recalled in Appendix~\ref{app:inversion}. 

 From \eqref{twopointnSDt} one can compute the infinite time limit of $G_i(x,t)$, denoted $G_i(x)$,
and of $D(L,t)$, denoted $D(L)$. These are still random quantities, exhibiting sample to sample fluctuations, and we can 
compute their equilibriumary distribution using the following identities:
\begin{align} 
    G_i(x) &= \langle  (h_{i+x}-h_i)^2 \rangle 
    \nonumber\\
    &= (\Lambda^{-1})_{i+x , i+x}+(\Lambda^{-1})_{i , i} -2 \, (\Lambda^{-1})_{i , i+x}\label{GLambda} 
    \\
    D(L)&= \frac{1}{L} {\rm Tr}\left( \Lambda^{-1} \right)
    = \frac{1}{L} \sum_{i=1}^L (\Lambda^{-1})_{i,i}  
\label{idGD}
\end{align}

\subsection{Free case}

Form the appendix \ref{app:inversion} we obtain the inverse of $\Lambda$ for free boundary conditions (see also Section~\ref{sec:boundary} for the construction of the matrix $\Lambda$),
and hence also the two point correlation of the interface in its equilibrium state as
\begin{equation} 
\label{Lambdafree}
  \langle h_i h_j \rangle  =  
  (\Lambda^{-1})_{ij}
  = \sum_{k=1}^{\min(i,j)} X_k
  \:.
\end{equation}
Here  $X_i = 1/k_i$ and the $X_i$'s are thus independent Pareto distributed random variables, with probability density $q(X)=  \mu\, X^{-1-\mu}$ for $X \in [1,\infty[$.

It is interesting to note that \eqref{Lambdafree} shows that for a given configuration of springs, equilibrium thermal fluctuations lead to an interface having the properties of a random walk for independent but \textit{not} identical increments:
i.e. we can represent the "free" interface at equilibrium (for $t\to\infty$) as $h_x  \eqlaw \sum_{i=1}^x \eta_i$ where $\eta_i$ are independent centered Gaussian random variables of variance $X_i=1/k_i$~; here $\eqlaw$ means equality in law relating two random quantities with same statistical properties.

As we will see, for $\mu<1$, sums of variables $X_k$ are dominated by their few largest elements, hence 
the interface will exhibit local jumps at the locations of these weak links.

To quantify this picture, it is convenient to introduce $\xi_x$, the sum of $x$ independent Pareto variables, distributed according to $\pi_x(\xi)$.  In this way, we can write  
\begin{align}
 \label{Gavg} 
     G^\free_i(x) &=  \sum_{j=i+1}^{i+x} X_j  \, \,
     \eqlaw \, \,  \xi_x
     \\
     \label{Davg}
    D^\free(L)&=  \sum_{j=1}^L \frac{L-j+1}{L} X_j 
\end{align}

We see that the
distribution of $G^\free_i(x)$ does not depend on $i$, and we denote it as
$P^\free_x(G)$. It thus identifies with $\pi_x(\xi)$, the distribution of the sum of $x$ Pareto  variables, $P^\free_x(G)=\pi_x(G)$, and has the following features: 
\begin{itemize}
\item[(i)] the support of $P^\free_x(G)$ is in the interval $(x,\infty)$ 
\item[(ii)]$P^\free_x(G)$  is peaked around the  $ G \sim x^{2 \zeta}$. For $\mu>1$, $2 \zeta=1$ (law of large numbers); for $\mu<1$, $2 \zeta=1/\mu$~;
in the marginal case $\mu=1$, $ G^\free(x) \sim x \ln x$.
\item[(iii)] For larger values, 
it decays as a power law $P^\free_x(G)\sim x G^{-\mu-1}$.
\end{itemize}

For $\mu>1$, the average $\overline{G^\free(x)}= \overline{\xi_x}=x\,{\mu}/{(\mu-1)}$ is finite. 
Hence $G{^\free}(x) \sim x$ and $\zeta=\zeta_\mathrm{EW}=1/2$.
On the contrary, for $\mu<1$, the stationary mean value $\overline{G^\free(x)}$ is infinite.
Moreover, at large $x$, the distribution of the scaled variable $\tilde G=G{^\free}(x)/[c_\mu x]^{1/\mu}$ 
is the one-sided Levy distribution of index $\mu$, denoted here $\mathcal{L}_{\mu,1}(\tilde G)$, and $c_\mu\sim\mathcal{O}(1)$ a dimensionless coefficient (see Appendix~\ref{app:distlevy}). The validity of the above results is tested numerically in Fig.~\ref{fig:distgasympt} for $\mu<1$. In Appendix~\ref{app:distmu1} we discuss also the case
$\mu >1$.

The equilibrium distribution $Q^\free_L(D)$ of the mean square displacement $D^\free(L)$ can also be computed, and has the following features: 
\begin{itemize}
\item[(i)] The support of $Q^\free_L(D)$ is in the interval $(L,\infty)$ 
\item[(ii)]$Q^\free_L(D)$  is peaked around $ D \sim L^{2 \zeta}$. 
\item[(iii)] For even larger values, it  decays as $Q^\free_L(D) \sim  L\,D^{-\mu-1}$  
\end{itemize}
(see detailed analysis in Appendix~\ref{app:distlevy}). 
Moreover, for $\mu<1$ at large $L$, the distribution of the scaled variable $\tilde D=D^\free(L)\big(\frac{1+\mu}{L c_\mu}\big)^{1/\mu}$ is the one-sided Levy distribution of index $\mu$, $\mathcal{L}_{\mu,1}(\tilde D)$ (see Eq.~\eqref{eq:DistribQfree} in Appendix \ref{app:distlevy}).
Hence, with free boundary condition the stationary mean value diverges for $\mu<1$, $\overline{D^\free(L)}=\infty$. This divergence is due to the presence of rare interfaces that contain atypical very large jumps.

\subsection{Fixed case}

From the appendix \ref{app:inversion} we obtain the inverse of $\Lambda$ for Dirichlet boundary conditions
(see Section~\ref{sec:boundary} for the definition of the matrix $\Lambda$ in this case)
and hence also the two point correlation of the interface in its equilibrium state as
\begin{equation}
  \langle h_i h_j \rangle  = 
    (\Lambda^{-1})_{ij}
= \frac{\left(\sum_{k=1}^i X_k \right)\left(\sum_{l=j+1}^{L+1} X_l\right)}{\sum_{p=1}^{L+1} X_p}
  \quad , \quad i \leq j 
\end{equation}
where, again $X_i=1/k_i$.
 Using \eqref{GLambda} we obtain
\begin{align} 
\label{eq:Gfixed}
G_i^\fixed(x) = \frac{(\sum_{k=i+1}^{i+x} X_k ) ( \sum_{p=1}^{L+1} X_p -\sum_{l=i+1}^{i+x} X_l) }{\sum_{p=1}^{L+1} X_p }
\end{align} 
Since the $X_k$ are independent we see that $G_i^\fixed(x)$ has a one site PDF which is independent of $i$ and is
distributed as 
\begin{align}
\label{Gavgp}
    G^\fixed(x) &  \eqlaw  
    \frac{\xi_x \tilde\xi_{L-x}}{\xi_x + \tilde\xi_{L-x}}
\end{align}
Here $\xi_x$ and $\tilde\xi_{L-x}$ are independent and distributed according to $\pi_x(\xi)$ and $\pi_{L-x}(\xi)$, respectively. 
One also finds

\begin{equation}
\label{eq:Dfixed}
D^\fixed(L) =  \frac{1}{L} \frac{\sum_{k=1}^L \sum_{l=k+1}^{L+1} (l-k) X_k X_l }
{\sum_{k=1}^{L+1}X_k}
\end{equation}

The distribution of $ G^\fixed(x) $ can then be written as:
 \begin{equation}
      P^\fixed_x(G) 
  =
 \int \int  \D \xi_1 \D \xi_2 \,\pi_x(\xi_1)\,\pi_{L-x}(\xi_2)\,\delta\left(G - \frac{\xi_1 \xi_2}{\xi_1 + \xi_2}\right)
 \label{inttail}
 \end{equation}
 Note that $ G^\fixed(x)  \leq \xi_x,\: \tilde\xi_{L-x} $. 
 To simplify the analysis, we discuss  the case where $x \ll L$, which allows to replace $\pi_{L-x}(\xi)$ by $\pi_{L}(\xi)$ in the integral. 
We distinguish two cases:
 \begin{itemize}
 \item 
   If $ G^\fixed(x) \ll L^{1/\mu}$ (or $\ll L$ if $\mu >1$)  we expect $ G^\fixed(x) \sim \xi_x$ like for the free case. 
 \item 
   If instead $ G^\fixed(x) > L^{1/\mu}$ (or $>L$ if $\mu >1$) then the integral \eqref{inttail} can be evaluated using the asymptotic power law decay  of both $\pi_x$ and $\pi_L$. 
 \end{itemize}
After some basic algebra, see Appendix \ref{app:distlevy}, one finds that the typical values of $G^\fixed(x)$ are around $\sim x^{1/\mu}$ for $\mu <1$ or around $\sim x$ for $\mu >1$.
For larger values the distribution of $G^\fixed(x)$ displays first a slow power law decay as $\sim x/G^{1+\mu}$, 
followed by a faster decay as $\sim xL/G^{1+2 \mu}$. 
The crossover thus occurs at for $G\sim L^{1/\mu}$. 
To summarize, we have 
\begin{align}
\label{asymptG}
P^\fixed_x(G) \sim
\begin{cases}
 x /G^{1+\mu} \quad & \text{for } \; x^{1/\mu} \ll G \ll L^{1/\mu}  
 \\[0.25cm]
 x L /G^{1+2 \mu} \quad & \text{for } \;  G \gg L^{1/\mu} .
\end{cases}
\end{align}
The validity of Eq.~\eqref{asymptG} is tested numerically in Fig.~\ref{fig:distgasympt} for $\mu<1$. 
In Appendix~\ref{app:distmu1} we discuss also the case
$\mu >1$.

Let us emphasize that the two different power laws originate from two different kinds of fluctuations: The first decay characterizes the values of $G^\fixed(x)$ obtained within a typical interface of size $L$. The second and faster decay characterizes the values of $G^\fixed(x)$ obtained only for disorder configurations that contain at least {\it two} atypically weak springs.

For $\mu>1$, the first power law, as in the free case, gives a finite  $\overline{G^\fixed(x)}$.
For $1/2<\mu<1$, in contrast with the free case, $\overline{G^\fixed(x)}$ is finite because of the faster decay of the second power law and the  scaling of Eq.\eqref{eq:anomscaling} is  recovered:
\begin{align}
    &\overline{G^\fixed(x)} 
    \\\nonumber
    &\sim x \int_{x^{\frac{1}{\mu}}}^{L^{1/\mu}} \D G\, 
    G\, \frac{1}{G^{1+ \mu}} 
    + x L \int_{L^{1/\mu}}^{+\infty} \D G\, 
    G\, \frac{1}{G^{1+ 2 \mu}} 
    \sim x L^{\frac{1}{\mu}-1}
    \:.
\end{align}

For $\mu<1/2$, the second term is divergent and $\overline{G^\fixed(x)}$ is infinite. 

The behaviour of the distribution of $D^\fixed(L)$ can also be derived: $Q^\fixed_L(D)$  is still peaked around   $ D \sim L^{2 \zeta}$, but the following power law decays as $\sim 1/D^{1+2\mu}$ (See Appendix~\ref{subsec:Dfixed}).
\\

We have thus characterized the fluctuations in the equilibrium regime. In Appendix D we indicate an
extension of these results in the case of a continuum model.
In the next section we study the problem at finite time, which requires information on the spectral density of the matrix ensemble~$\Lambda$.


\section{Finite time Regime}
\label{sec:FTR}

\begin{figure}[!ht]
\centering
\includegraphics[width=0.45\textwidth]{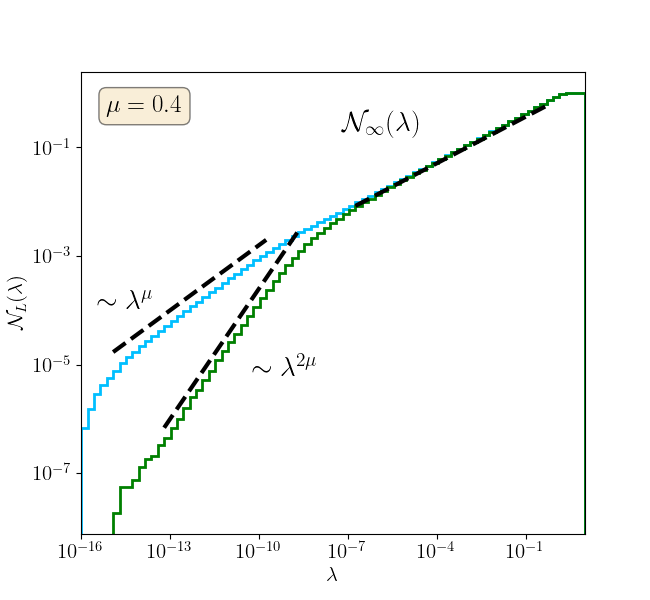}
\includegraphics[width=0.45\textwidth]{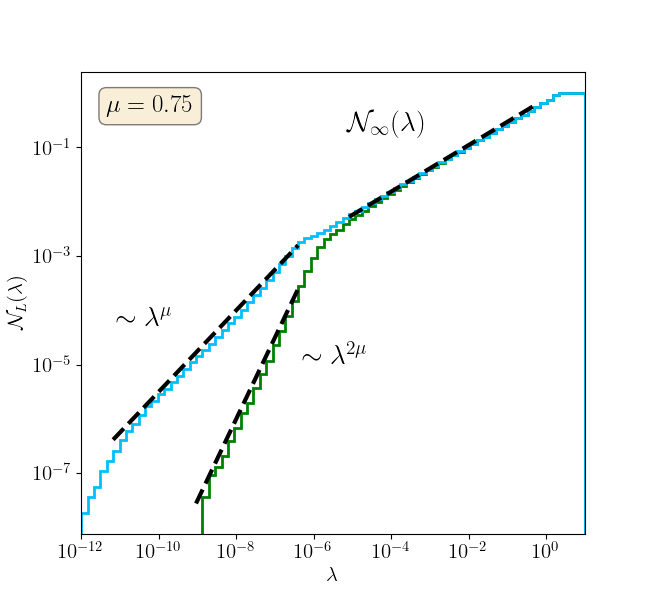}
    \caption{Cumulative PDF of the eigenvalues $\mathcal{N}_L (\lambda)$ for $\mu=0.4$ and $N_s=10^8$ realisations (top) and for $\mu = 0.75$ (bottom) for $L=500$ for free case (blue) and free case (green). The black dashed line corresponds to the analytical predictions: the predicted asymptotics and $\mathcal{N}_\infty(\lambda) =\int_0^\lambda \rho(s) \D s$.} 
    \label{DoS2}
\end{figure}

To compute two-points correlation functions at finite time, we use the explicit expression given in Eq. \eqref{twopointsSDt}. Hence, the mean-square displacement display a particular simple form: 
\be
\overline{D(L,t)} = \frac{1}{L} \overline{\text{Tr}\left(\frac{1 - e^{-2 \Lambda t}}{\Lambda}\right)} = \frac{1}{L}  \overline{\sum_{\alpha=1}^L \frac{1 - e^{-2 \lambda_\alpha t}}{\lambda_\alpha}}.
\ee
Indeed, the right-hand size of the equation depends only on the eigenvalues $\lambda_\alpha$ of $\Lambda$ because the trace is basis invariant. 
The average over the disorder realisations can be expressed in terms of the average spectral density 
$\rho_L(\lambda)= \frac{1}{L} \overline{\sum_\alpha \delta(\lambda- \lambda_\alpha)}$ of the ensemble of the matrices~$\Lambda$.
\be
\overline{D(L,t) } = \int \D\lambda\, \rho_L(\lambda)\, \frac{1 - e^{-2 \lambda t}}{\lambda} .  \label{Dt} 
\ee
We also introduce the notation $\rho(\lambda)=\lim_{L\to\infty}\rho_L(\lambda)$.
The expression for $\overline{G(x,t) }$ is not basis invariant and will be discussed later.

\subsection{Average spectral density and scaling of \texorpdfstring{$\overline{D(L,t)}$}{TEXT}}
The average spectral density $\rho_L(\lambda)$ is a function with positive support ($\Lambda$ is a positive matrix) and its behaviour close to the origin dominates the long time behaviour. 
For $\mu>1$, the spectral density presents, for $\lambda\to0$, the same behaviour as the non disorder case, namely $\rho(\lambda)\simeq1/\big[2\pi\sqrt{k_a\lambda}\big]$ provided that we use an average spring constant $1/k_a=\overline{1/k_i}$.
In Fig.~\ref{DoS2} we show the results for the cumulative distribution $\mathcal{N}_L(\lambda) =\int_0^\lambda \rho_L(s) \D s$ for $\mu<1$ (when $\overline{1/k_i}=\infty$). 
These results are compared with the exact asymptotic behaviour (valid when $L\to \infty$) which we have obtained by using the relation of the random spring chain with the analysis of certain random products of $2\times2$ matrices along the lines of \cite{Tex20}~:
\begin{equation}
\label{BernasconiDoS}
\rho(\lambda) \underset{\lambda \to 0}{\simeq}
\frac{\mu\, \Gamma(1-\mu)^{\frac{1}{1+\mu}}}{(1+\mu)^{\frac{2}{1+\mu}} \Gamma(\frac{1}{1+\mu})^2}
\:\lambda^{-\frac{1}{1+\mu}}  
\quad \text{for} \quad \mu<1.
\end{equation}
We will present in a forthcoming paper a derivation of this asymptotics. 
The exponent and the constant were first obtained in Ref.~\cite{stephen1982diffusion} by a different approach (See Eq.3.26 of this paper; this was obtained even earlier under a different form in  Ref.~\cite{bernasconi1980diffusion}).  
Here we observe the perfect agreement with the numerics, except at very small $\lambda$ where finite size corrections and the dependence on the boundary conditions  become important. Indeed a matrix $\Lambda$ has $L$ eigenvalues $\lambda_1<\lambda_2<\cdots<\lambda_L$ and the minimum among them has a typical value $\lambdaMinTyp$  given by the extreme value statistics:
\begin{equation}
    \mathcal{N}_L (\lambdaMinTyp)=1/L, \quad \; \lambdaMinTyp\sim L^{-(1+\mu)/\mu} 
\end{equation}
Hence, for $\lambda > \lambdaMinTyp$ we expect $\rho_L(\lambda)\approx \rho(\lambda)$ independently on the boundary conditions. From Eq.\eqref{twopointsSDt} we observe that eigenvalues are rates and the time scale associated to $\lambdaMinTyp$
is $\sim L^{-z}$ because $z=(1+\mu)/\mu$. Hence, the anomalous dynamical exponent $z$ can be derived from the behaviour of $\rho(\lambda)$ close to the origin. For $t \ll L^z$, using $\rho_L(\lambda) \simeq \rho(\lambda)$, we find that $\overline{D(L,t)}=\overline{D(t)}$ is independent of $L$, and recover also the exponent $2 \beta=1/(\mu+1)$: 
\begin{align}
\overline{D(t)}&=\int_0^{\infty} \D\lambda\, \rho(\lambda)\, \frac{1 - e^{-2 \lambda t}}{\lambda} \simeq  C_\mu  t^{2 \beta} , \quad  \text{for} \,t \ll L^z\\
C_\mu&=
\left(
  \frac{ 2\, \Gamma(1-\mu) }{ (1+\mu)^\mu }
\right)^{\frac{1}{1+\mu}}
\frac{\mu \, \Gamma(\frac{\mu}{1+\mu}) } {\Gamma(\frac{1}{1+\mu})^2} 
\label{comptempG}
\end{align}

Only few realizations of $\Lambda$ have eigenvalues   $\lambda <\lambdaMinTyp$. These rare configurations are characterized by a single very weak spring (free case) or  two very weak springs (fixed case) as shown in Fig.~\ref{figintv2}. The value of this (or these two) spring constant(s) should be smaller than $L^{-z} \sim L^{-1-1/\mu}$, well below the typical value $\sim L^{-1/\mu}$ of the weakest spring constant. In this limit, we have $\lambda_{1} \sim k_{\min}$ for the free case or equal to the second weakest spring for the fixed case. \\
Hence, for $\lambda <\lambdaMinTyp$, $\rho_L(\lambda) $ scale as $f_{1,L}(\lambda)$ for the free case or $f_{2,L}(\lambda)$ for the fixed case. By matching the small $\lambda$ behaviour to $\rho(\lambda)$ when $\lambda \sim \lambdaMinTyp$, we have

\begin{align}
    \label{rhoper}
    &\rho^\free_L(\lambda) \simeq
    \begin{cases}
     C^{\free } L^{\mu}  \lambda^{\mu-1} \quad &\text{for} \, \lambda< L^{-z}   
     \\[0.25cm]
     \rho(\lambda) \quad &\text{for} \, \lambda >L^{-z}  
    \end{cases}
\\[0.5cm]
     \label{rhofix}
  &  \rho^\fixed_L(\lambda) \simeq
    \begin{cases}
     C^{\fixed } L^{1+2\mu}  \lambda^{2\mu-1}  \quad &\text{for} \, \lambda< L^{-z}   
     \\[0.25cm]
     \rho(\lambda) \quad &\text{for} \, \lambda >L^{-z}   
     \end{cases}
\end{align}
Here, $C^\fixed$ and $C^\free$ are constants of order $1$.

 Injecting this result in \eqref{Dt} and performing the integral over $\lambda$, we recover the expression of $\overline {D^\free(L,t)}$ given in Eq. \eqref{scaleDmu05free} and $\overline{D^\fixed(L,t)}$ in Eq. \eqref{scaleDmu05per}.

\subsection{Scaling of \texorpdfstring{$\overline{G(x,t)}$}{TEXT}}

\begin{figure}[!ht]
\includegraphics[width=0.5\textwidth]{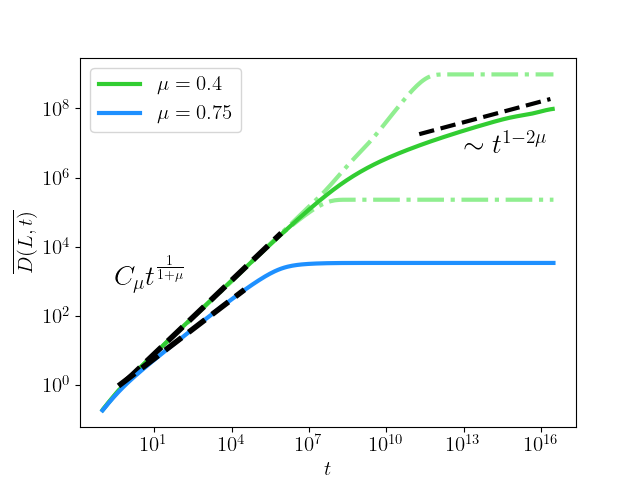}
\includegraphics[width=0.5\textwidth]{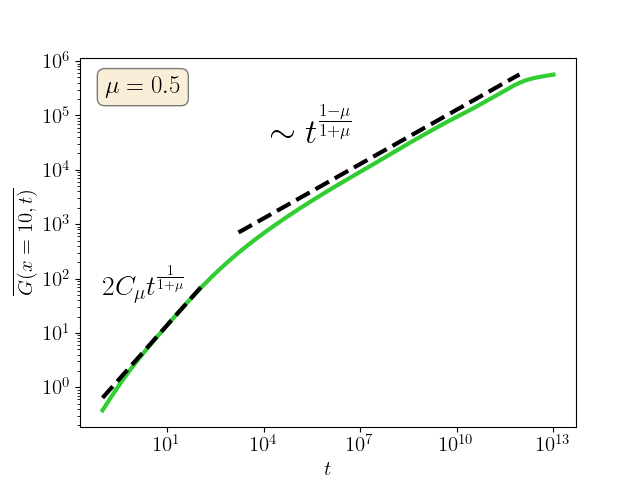}
    \caption{ Time dependent scaling in the fixed case. Top: $\overline{D(L,t)}$ for  $N_s=10^5$,$L=500$ with   $\mu=0.75$ (plain line Blue) and $\mu=0.4$ (plain line Green). The dashed lines correspond to $D(L,t)$ for two different disorder realisations. 
    Bottom: Time evolution of $\overline{G(x=10,t)}$ for $\mu=0.5$, $L=5000$, $N_s=10^6$.}
    \label{DtGx3}
\end{figure}

\begin{figure}[!ht]
\centering
\includegraphics[width=0.5\textwidth]{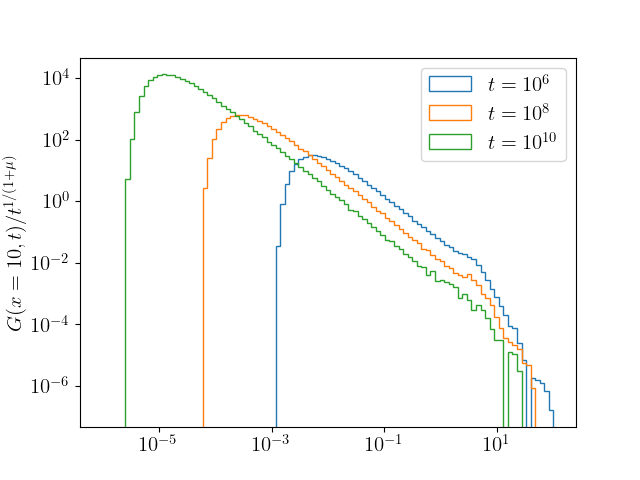}
    \caption{Rescaled histogram of $G(x=10,t)$ at various times for $\mu=0.4$, $L=5000$ and $N_s=10^6$ realisations. The realisations were obtained by sampling the spring constants and direct evaluation of Eq.\eqref{twopointsSDt} and Eq.\eqref{GLambda}. The cutoff of the power law occurs in the same region for the three histograms, which therefore scales as $G_{\max}(x,t) \sim t^{2 \beta}$.  
    }
    \label{Gx10EVTmu05}
\end{figure}

Here we study the  height-to-height correlation function at  finite time:
\begin{align}
    G_i(x,t) &= \langle (h_{i+x}(t) - h_i(t))^2 \rangle 
    \nonumber\\
    &= \langle h_{i+x}(t)^2 \rangle + \langle h_i(t)^2 \rangle - 2 \langle h_{i+x}(t)   h_i(t) \rangle
\end{align}
we focus again in the behaviour far from the boundary, where the dependence on $i$ can be neglected  and the heights   $h_{i+x}(t)$ and $h_i(t)$ are identical random variables.
For $\mu<1$, we distinguish three regimes for the scaling of $\overline{G(x,t)}$:
\begin{itemize}
    \item Short time, $t \ll x^z$. The growing length $\ell(t)$ is smaller than the window size $x$, such that the heights $h_{i+x}(t)$ and $h_i(t)$ are still not correlated. Their mean is zero  and their variance, $\sim \ell^{2 \zeta}(t) \sim t^{2\beta}$. So we can deduce [and retrieve Eq.\eqref{eq:anomscaling}]:
\[
\overline{G(x,t)} \approx 2 \,\overline{D(L,t)} \sim t^{2 \beta}
\] 
\item Intermediate time, $x^z \ll t \ll L^z$.  In this regime, the mean value of the height to height correlation function is dominated by the  largest jumps of the growing interface. Hence, the distribution of $G(x,t)$ mirrors the first power law decay of the equilibrium  distribution, namely $P_x(G) \sim x\,G^{-1-\mu}$ , albeit with a time-dependent cutoff at $G_{\max}(x,t) \sim  t^{2\beta}$. This is confirmed numerically Fig.~\ref{Gx10EVTmu05}, hence  we deduce: 
\begin{equation}
\overline{G(x,t)} \sim \int_{x^{{1}/{\mu}}}^{t^{2\beta}} \D G \, G \, \frac{x}{G^{1+ \mu}} \sim x t^{\frac{1-\mu}{1+\mu}}
\end{equation}
We thus retrieve Eq.\eqref{eq:anomscaling} and the so called "anomalous" scaling behaviour.
\item Large time, $L^z \ll t$. For $1/2 < \mu < 1$ and Dirichlet boundary conditions, $\overline{G(x,t)}$ reaches an equilibrium limit, in all other cases, it is dominated by the largest jumps among the interfaces that haven't equilibrated yet. We can simply infer the late time behaviour using the exact relation:
\begin{equation}
\overline{D(L,t)} =  \frac{1}{L} \sum_{x=1}^L\overline{ G(x,t)},
\end{equation}
and assuming the following scaling form:
\begin{equation}
\overline{G(x,t)} \sim x t^{\alpha_1} L^{\alpha_2},
\end{equation}
the  factor $\sim x$ arises from the probabilistic interpretation while the value of the exponents $\alpha_1$ and $\alpha_2$ depend on the boundary conditions. 
For Dirichlet boundary conditions we know that $\overline{D^\fixed}(L,t) \sim t^{1-2\mu} L^{1+2\mu}$, such that  $\alpha_1 = 1 - 2\mu$ and  $\alpha_2 = 2\mu$. This corresponds to the results of Eq.~\eqref{scalemu05per} and it is checked numerically in Fig. \ref{DtGx3}. Using a similar reasoning, one obtains the "free" case  and retrieves the scaling relation as given in Eq.~\eqref{eq:anomscaling}.
\end{itemize}

\section{Discussion}
\label{sec:lopez}

In this paper we discuss a variant of the Edward-Wilkinson interface in which the elastic constants of the  springs are non homogeneous and drawn from  the distribution (\ref{distspring}). The scaling properties of the interface are affected the behaviour of the distribution close to the origin $p(k) \sim k^{\mu-1}$ for $k \to 0$. 

In particular, for $\mu>1$ we recover the non disordered Edward-Wilkinson scaling with a homogeneous elastic constant $k_a$ such that $1/k_a=\overline{1/k_i}$. 

More interesting is the case  $\mu<1$, where $\overline{1/k_i}=\infty$. In this case, 
the scaling properties are characterized by new exponents and anomalous scaling. This new behaviour was first  predicted in \cite{lopez1996lack,lopez1997superroughening,lopez1999scaling,lopez1997power}  through renormalization group techniques. 
Here, we revisited this problem using (i) analytic representations of the equilibrium correlation of the line for any given realization of the random springs, 
(ii) results for the spectral properties of the random matrix $\Lambda$, obtained using techniques developed for products of random $2\times2$ matrices \cite{Tex20} (details on this second point will be given in a forthcoming paper).
We recover the same exponents obtained in previous studies, but we  disagree with the anomalous scaling in two points:
\begin{itemize}
    \item  At intermediate and large times, we note that $\overline{G(x,t)}$ grows as $ \sim x $ (see Eqs.(\ref{eq:anomscaling}) and (\ref{scalemu05per}). Lopez {\it et al.}  provide  a geometrical interpretation of this linear dependence \cite{lopez1996lack,lopez1997superroughening,lopez1999scaling,lopez1997power}. Indeed, they write  $\overline{G(x,t)} \sim   x^{2\zeta_{\text{loc}}}$, with $\zeta_{\text{loc}}=1/2$.
 The exponent $\zeta_{\text{loc}}$ is supposed to be a "local" roughness exponent $\zeta_{\text{loc}}=1/2$ that characterizes the local properties of the interface, in contrast with the "global" roughness exponent $\zeta=1/(2\mu)$. We disagree with this picture and provide a probabilistic interpretation: the  typical values for $G(x,t)$ scale as $x^{2 \zeta}$. On the other hand, the mean value,  $\overline{G(x,t)}$, is dominated by the rare pieces of interface that contain the largest jump. Hence, the factor $x$ is originated by the  probability of such event.
  \item Using our results for the distribution $P_x(G)$ (Section \ref{sec1} and \ref{sec:FTR}), we can also analyze the multiscaling properties. We get $\overline{G(x,t)^{q/2}} \sim x^{\zeta_{\text{loc}}(q) q}$, which can be simply understood ($\zeta_{\text{loc}}$ introduced above is $\zeta_{\text{loc}}(q=2)$)~: for large moments, $q>q_c$ the average is dominated by the presence of these abrupt jumps. Hence, we have $\zeta_{\text{loc}}(q) = 1/q$. For small enough moments $q<q_c$, instead, these jumps become irrelevant and the average coincide with the typical behaviour, hence $\zeta_{\text{loc}}(q) = \zeta$. By continuity, we then recover $q_c = 1/ \zeta$ and the multiscaling can be described as:
\be
 \zeta_{\text{loc}}(q) =
    \begin{cases}
 \quad \zeta \quad & \text{for } \;  q < 1/\zeta   
    \\[0.25cm]
 \quad 1/q  \quad & \text{for } \;  q > 1/\zeta
\end{cases}
\ee
 \item Lopez {\it et al.} predict that at large times, $\overline{D(L,t)} \sim L^{2 \zeta}$ and $\overline{G(x,t)} \sim x L^{2 \zeta -1}$ for $t \to \infty$, independently of the boundary conditions. We disagree and we have shown that rare disorder realization with one or two very weak spring completely modify this result.
\end{itemize}

When $\mu<1$ and $t\gg L^z$, we find that the average value of the observable crucially depends on the boundary conditions. We discussed in detail two cases: (i) the free case where  an  extremity is attached at the origin and the second is free and (ii) the fixed case in which both extremities are attached at the origin. In both cases the center of mass of the line is confined and the observable $D(L,t)$ probes the fluctuations of the span of the interface. 
For periodic boundary conditions or for free boundary conditions on both extremities,
two cases not studied here, the center of mass undergoes unbounded free diffusion. 
If one subtracts this zero mode motion, we expect that the periodic boundary conditions displays the same behaviour as the one for the fixed case, and the line with two free extremities behaves as the one with a single free boundary. 

As clearly stated in the introduction our results hold in the bulk of the line, far from the extremities. Close to the extremities we observed numerically that the behaviour of $G_i(x,t)$ is  different from the bulk and crucially depends on the particular prescription (free or Dirichlet). We plan to study this behaviour in a future work.


\section*{Acknowledgements}
We thank A. Fedorenko for discussions on related topics. 
PLD acknowledges support from 
ANR grant ANR-23-CE30-0020-01 EDIPS,
and thanks KITP for hospitality, 
supported in part by the National Science Foundation Grant No. NSF PHY-1748958 and PHY-2309135.


\begin{appendix}

\section{Matrix inversion} \label{app:inversion}

We explain how to invert the $L \times L$ tridiagonal and symmetric matrix $\Lambda$. Here, the diagonal elements  are $\Lambda_{i,i}=k_i + k_{i+1}$, and the off-diagonal elements $\Lambda_{i,i+1}=\Lambda_{i+1,i}= -k_{i+1}$.  
The aim is to find the matrix $C_{ij}=(\Lambda^{-1})_{ij}$ such that 
\begin{equation}
  \label{eq:InversionLambda}
    \sum_{l=1}^L \Lambda_{il}C_{lj}=\delta_{ij}
    \:.
\end{equation}
We follow the same strategy as for the construction of the Green's function of a second order linear operator.

\subsection{Dirichlet/Dirichlet}
We consider first the Dirichlet boundary conditions ("fixed case"), although the method applies to any choice of boundary conditions.
We introduce the solution of $\Lambda\psi=0$ which satisfies the left boundary condition, i.e. $\psi_0=0$, hence $\psi_i$ can be obtained from the recurrence
\begin{equation}
  \label{eq:reccu1}
  \psi_{i+1} = \left( 1 + \frac{k_{i}}{k_{i+1}} \right) \psi_i - \frac{k_{i}}{k_{i+1}}  \psi_{i-1} 
  \:.
\end{equation}
Furthermore, for convenience, we choose the solution such that $\psi_1=1/k_1$. As a result
\begin{equation}
    \psi_i = \sum_{n=1}^i\frac{1}{k_n}
    \:.
\end{equation}
Similarly, we introduce the solution of $\Lambda\chi=0$ which satisfies the right boundary condition $\chi_{L+1}=0$, with $\chi_L=1/k_{L+1}$ for convenience.
Solving the recurrence 
\begin{equation}
  \label{eq:reccu2}
  \chi_{i-1} =  \left( 1 + \frac{k_{i+1}}{k_{i}} \right) \chi_i - \frac{k_{i+1}}{k_{i}}  \chi_{i+1} 
\end{equation}
we obtain 
\begin{equation}
  \chi_i = \sum_{n=i+1}^{L+1} \frac{1}{k_n}
  \:.
\end{equation}
The inverse of $\Lambda$ can thus be expressed in terms of the two solutions as 
\begin{equation}
  C_{i,j} = A \, 
  \begin{cases}
    \psi_i \, \chi_j
   \hspace{0.5cm} \mbox{for } i\leq j
   \\
    \psi_j \, \chi_i
   \hspace{0.5cm} \mbox{for } i\geq j
   \end{cases}
\end{equation}

In order to find the constant $A$ we write \eqref{eq:InversionLambda} for $i=j$
\begin{equation}
   A\left(  -k_{j}\psi_{j-1}\chi_{j}
    +(k_{j}+k_{j+1}) \psi_{j}\chi_{j}
    -k_{j+1}\psi_{j}\chi_{j+1} 
   \right)
    =1
\end{equation}
Using the expressions of the two solutions, we deduce that 
$1/A=\sum_{j=1}^{L+1}k_j^{-1}$, i.e.

\begin{align}
\label{eq:InvLambda}
&    (\Lambda^{-1})_{ij}= \frac{\left(\sum_{k=1}^i X_k \right)\left(\sum_{l=j+1}^{L+1} X_l\right)}{\sum_{p=1}^{L+1} X_p}  \quad , \quad i \leq j 
\\[0.25cm]
&  (\Lambda^{-1})_{ij}= (\Lambda^{-1})_{ji}  
\quad , \hspace{3cm} i > j 
\end{align} 
with  $X_i = 1/k_i$.

The method can be applied to any tridiagonal matrix. 
Here it is particularly simple as the constraint 
$\Lambda_{i,i-1}+\Lambda_{i,i}+\Lambda_{i,i+1}=0$ has allowed to solve analytically the recurrences (\ref{eq:reccu1},\ref{eq:reccu2}) and get simple forms for the two independent solutions $\psi_i$ and $\chi_i$, which is not always possible for any tridiagonal matrix.

A similar discussion can be found in Ref.~\cite{usmani1994inversion}.

\subsection{Dirichlet/Neumann}

 In the case of free boundary condition for $h_L$, one takes the limit $k_{L+1} \to 0$ in Eq.~\eqref{eq:InvLambda}, and one obtains 
Eq. \eqref{Lambdafree} in the text.


\section{Distribution of \texorpdfstring{$G(x)$}{TEXT} for \texorpdfstring{$\mu<1$}{TEXT}} \label{app:distlevy}

The inverse of the spring constants $X_i=1/k_i$ play a central role. They are i.i.d. random variables obeying a Pareto distribution $q(X)=\mu\,X^{-1-\mu}$ for $X>1$, with $\mu>0$.
Its Laplace transform has a simple form 
\begin{equation}
  \label{eq:LaplacePareto}
    \tilde{q}(s) = \int\D X\,q(X)\,e^{-sX}
    =e^{-s} - s^\mu\,\Gamma(1-\mu,s)
\end{equation}
where $\Gamma(\alpha,s)=\int_s^\infty\D t\,t^{\alpha-1}\,e^{-t}$ is the incomplete Gamma function.
Note that \eqref{eq:LaplacePareto} also holds for $\mu>1$ (next appendix).

\subsection{Distribution of \texorpdfstring{$G(x)$}{TEXT} in the free case}

The Laplace transform of the PDF $\pi_n(\xi)$ of $\xi_n=\sum_{j=1}^n X_j$, 
 is obviously
\begin{align}   
\label{LaplacePi} 
  \tilde\pi_n(s)
  =
 \int \D\xi\, \pi_n(\xi)\, e^{-s \xi} 
 &= \tilde{q}(s)^n
\end{align}   
 For $0< \mu<1 $, it is clear from \eqref{eq:LaplacePareto} that 
 $\tilde{q}(s) \simeq 1 - \Gamma(1-\mu) \, s^\mu+ \mathcal{O}(s)$, hence
 $\tilde\pi_n(s)\simeq\exp\{-\Gamma(1-\mu)\,n\,s^\mu\}$ at large $n$ and small $s$ with $n\,s^\mu\sim\mathcal{O}(1)$.
 We deduce that the large $n$ form of  the PDF of $\xi_n$ for $\mu<1$ is, upon rescaling, a one-sided ($\beta=1$) Levy stable distribution of index $\mu$. 
 
We recall that, {\bf more generally,} the L\'evy distribution is defined as 
\begin{align}
\label{eq:LevyDistribution}
 & \mathcal{L}_{\mu,\beta}(x)
  = \int_{\mathbb{R}} \frac{\D k}{2 \pi} e^{\I k x} 
  e^{-|k|^\mu
    \big[ 
      1 +\I\, \beta\, {\rm sgn}(k) 
      \tan( \frac{\pi \mu}{2}) 
  \big] } 
  \\\nonumber
&\hspace{0.5cm}
 \mbox{for } \mu\in]0,1[\cup]1,2]
\end{align}
(see e.g. Appendix B in \cite{bouchaud1990anomalous} and Appendix A in \cite{bouchaud1990classical} and references therein, or the recent monograph \cite{App04}). 
$\beta\in[-1,+1]$ is an asymmetry parameter.
In particular, the asymptotic behaviour is 
\begin{equation}
 \label{eq:AsymptoticLevy}
  \mathcal{L}_{\mu,\beta}(x) \underset{ x\to\pm\infty }{\simeq}  \frac{\sin(\frac{\pi\mu}{2})\,\Gamma(\mu+1)}{\pi}\,
    \frac{1+\beta\,\mathrm{sgn}(x)}{|x|^{\mu+1}}
    \:,
\end{equation}
hence for $\beta=1$, the power law exists only on $\mathbb{R}^+$.
For $0<\mu<1$ and $\beta=1$ the L\'evy distribution is defined for $x>0$.
For $\mu=1/2$ it takes the simple form  
\begin{equation}
  \mathcal{L}_{\frac12,1}(x)= \frac1{ \sqrt{2\pi} x^{3/2}}e^{-1/(2 x)}
\end{equation}
and in general it has an essential singularity at the origin $\mathcal{L}_{\mu,1}(x) \sim \exp\big\{-\gamma_\mu\,x^{-\mu/(1-\mu)}\big\}$ for $x\to0$.
Other formulas and properties can be found in e.g. \cite{penson2010exact}. 

When $\beta=1$, which is the case of interest here, the Fourier trasnform of the L\'evy distribution is $\hat{\mathcal{L}}_{\mu,1}(k)=\exp[-(\I k)^\mu/\cos(\pi\mu/2)]$, so that the above behaviour is 
$\tilde\pi_n(\I k)\simeq\hat{\mathcal{L}}_{\mu,1}\big((c_\mu n)^{1/\mu}k\big)$, 
where $c_\mu=\Gamma(1-\mu)\cos(\pi\mu/2)$. 
As a result 
\begin{equation}
  \label{eq:PiMuSmall}
    \pi_n(\xi) \simeq \frac{1}{(c_\mu n)^{1/\mu} } 
\:\mathcal{L}_{\mu,1}\left(\frac{\xi}{(c_\mu n)^{1/\mu}}\right)
\end{equation}
The asymptotic of $\pi_n(\xi)$ is useful in the paper~:
we deduce from \eqref{eq:AsymptoticLevy} that $\pi_n(\xi) \simeq (\mu n)/\xi^{1+\mu}$ at large $\xi$,
as stated in the main text.

It is useful to compare $\pi_n(\xi)$ with the PDF of $X_{\rm max}$ the largest of the $n$ variables $X_i$, which is, from \eqref{weakest}
\begin{equation}
f_{1,n}(X)= \frac{\mu \,n}{X^{1+\mu} }e^{-n X^{-\mu}}
\end{equation}
It has the same tail as $\pi_n(\xi)$, i.e. $\xi_n \simeq X_{\rm max}$ in the tail, while
they are of the same order $\xi_n \sim X_{\rm max} \sim n^{1/\mu}$ for typical events.

\subsection{Distribution of \texorpdfstring{$D(L)$}{TEXT} in the free case}

Consider now the PDF of $\theta_L = \sum_{j=1}^L a_j X_j$, where $a_j$ are some $\mathcal{O}(1)$ coefficients. 
The same 
manipulations as above show that its Laplace transform is now 
\begin{equation}
    \langle e^{-s\theta_L}\rangle=\prod_{j=1}^L\tilde{q}(a_j s)\simeq
    \exp\bigg\{ 
      - \Gamma(1-\mu) \,s^\mu \sum_{j=1}^L a_j^\mu 
    \bigg\}
\end{equation}
at small $s$ with $s^\mu L\sim\mathcal{O}(1)$. 
Hence it is now the scaled random variable
\begin{equation}
\tilde \theta 
= \frac{\theta_L}{\left( c_\mu \sum_{j=1}^L a_j^\mu\right)^{1/\mu}}
\end{equation}
which is distributed with a one-sided Levy stable distribution of index $\mu$, $\mathcal{L}_{\mu,1}(\tilde \theta )$. 
In the text $a_j=(L-j+1)/L$ and $\theta_L\to D^\free(L)$.  
For large $L$ one has
\begin{equation} 
  \sum_{j=1}^L a_j^\mu \simeq L \int_0^1 (1-x)^\mu \D x = \frac{ L}{1+\mu} 
\end{equation}

hence, at large $L$,
\begin{equation}
  \label{eq:DistribQfree}
  Q^\free_L (D) 
  \simeq 
  \left(\frac{1+\mu}{L \,c_\mu } \right)^{1/\mu}
  \:
   \mathcal{L}_{\mu,1}\left(\left(\frac{1+\mu}{L \,c_\mu } \right)^{1/\mu} D \right)
  \:.
\end{equation}

\subsection{Distribution of \texorpdfstring{$G(x)$}{TEXT} in the fixed case}

Consider now 
\begin{equation} 
  \frac{1}{ G^\fixed(x)} = \frac{1}{\xi_{x}} + \frac{1}{ \tilde\xi_{L-x}} 
  \:.
\end{equation}
We define 
 the two independent variables $\zeta_x=1/\xi_{x}$ and
$\tilde\zeta_{L-x}=1/ \tilde\xi_{L-x}$ with 
PDF 
$Q_x(\zeta)=\zeta^{-2} \pi_x(1/\zeta)$
and $Q_{L-x}(\zeta)$.
For $\mu<1$ it behaves as $\sim\mu x \zeta^{\mu-1}$ at small $\zeta$.
The PDF of $U=1/G=\zeta_x+\tilde\zeta_{L-x}$ is obtained by a simple convolution, and its behaviour for $U\to0$ is 
\begin{align} 
 U^{-2}\,P_x^\fixed(U^{-1}) &=
\int_0^U \D \zeta \, Q_{x}(\zeta)\, Q_{L-x}(U-\zeta) 
\nonumber\\
&
\underset{U \to 0}{\simeq} \mu^2 x (L-x) \int_0^U \D\zeta \, \zeta^{\mu-1} (U-\zeta)^{\mu-1}  
\nonumber\\
&
= \frac{\Gamma(1+\mu)^2}{\Gamma(2 \mu)} x(L-x) U^{2 \mu-1}
\end{align}

which upon inversion leads to the tail of $ P_x^\fixed(G)$ at large $G$
\begin{equation} 
 P_x^\fixed(G) \simeq \frac{\Gamma(1+\mu)^2}{\Gamma(2 \mu)} \frac{x(L-x) }{G^{1 + 2 \mu} }
\hspace{1cm}\mbox{for }
G\to\infty
\end{equation}

\subsection{Distribution of \texorpdfstring{$D(L)$}{TEXT} in the fixed case}
\label{subsec:Dfixed}

\begin{figure}[!ht]
\centering
\includegraphics[width=0.5\textwidth]{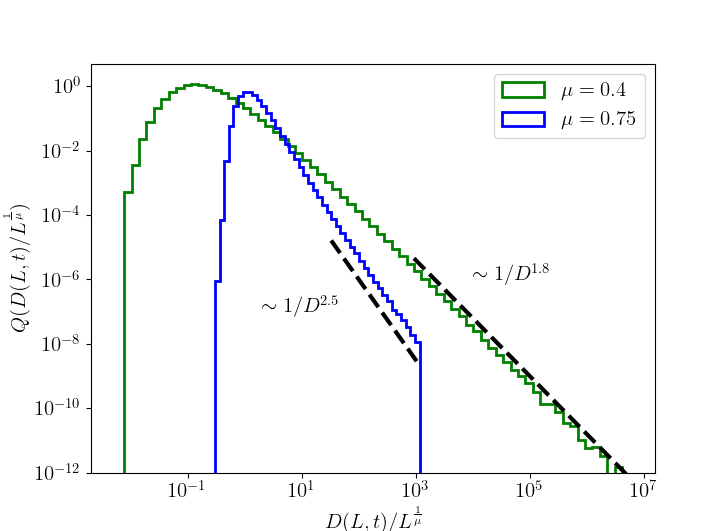}
    \caption{
    Histogram of $D(L)\equiv D(L,t\to\infty)$ in the equilibrated regime for $\mu=0.4,0.75$ and $L=5000$ for the fixed case. $N_s=10^7$ values of $D(L)$ is  obtained by direct evaluation of  Eq. \eqref{Davg}, 
    For each distribution, the typical behaviour is located around $G \sim L^{1/\mu}$ ($2 \zeta=1/\mu$). The power law decay is given by $D(L) \sim L^2/ D ^{1+ 2\mu}$.
     }
    \label{fig:distDasympt}
\end{figure}

The two sums \eqref{eq:Gfixed} and \eqref{eq:Dfixed} have the same structure. We do not expect the additional factors $(l-k)/L$ to affect the scaling properties.
The two main properties for $P_x^\fixed(G)$, distribution peaked at $x^{1/\mu}$ and tail $\sim G^{-1 - 2 \mu}$, translate as $Q_L^\fixed(D)$ peaked at $L^{1/\mu}$ and tail $\sim D^{-1-2\mu}$ [the correspondence between $P_x^\fixed(G)$ and $Q_L^\fixed(D)$ requires $x\sim L$, hence there is no intermediate regime as in Eq.~\eqref{asymptG}].
We have checked this numerically in  Fig.~\ref{fig:distDasympt}.

\section{Distribution of \texorpdfstring{$G(x)$}{TEXT} for \texorpdfstring{$\mu\geq1$}{TEXT}} \label{app:distmu1}

\subsection{Summary of the results for \texorpdfstring{$\mu>1$}{TEXT}}

For $\mu>1$, the power law are identical to the case $\mu<1$, although position of the peaks and crossover are differents. In the free and fixed cases, the distribution are peaked around their mean, ie. $G \sim x$ and $D \sim L$.
As a consequence, in the fixed case, the crossover of $G$ now occurs at $L$:

\begin{align}
\label{asymptGmusup1}
P^\fixed_x(G) \sim
\begin{cases}
 x /G^{1+\mu} \quad & \text{for } \; x \ll G \ll L
 \\[0.25cm]
 x L /G^{1+2 \mu} \quad & \text{for } \;  G \gg L .
\end{cases}
\end{align}

\subsection{Distribution of \texorpdfstring{$G(x)$}{TEXT} in the free case} 

Expression \eqref{eq:LaplacePareto} also holds for $\mu>1$, however the incomplete Gamma function is now divergent in the limit $s\to0$~: we should use $\Gamma(-q,s)\simeq (1/q)\,s^{-q}+\Gamma(-q)+\mathcal{O}(s)$ for $q\in]0,1[$,
hence
\begin{equation}
    \tilde q(s) \underset{s\to0}{ \simeq }
    1 - \frac{\mu}{1-\mu}\,s 
    - \Gamma(1-\mu)\,s^\mu+\mathcal{O}(s^{2})
    \hspace{0.5cm}\mbox{for }1<\mu<2
    \:.
\end{equation}
The first term is obviously the contribution of $\overline{X_i}=\mu/(\mu-1)$.

It is more simple to relate the Laplace transform to the Fourier transform and write
\begin{equation}
   \pi_n(\xi)
   =\int\frac{\D k}{2\pi} \, \tilde q(\I k)^n\,e^{\I k\xi}
   \simeq
   \int\frac{\D k}{2\pi} \, 
   e^{\I k\xi -\I k \overline{\xi_n} - n\, \Gamma(1-\mu)\,(\I k)^\mu}
\end{equation}
where $\overline{\xi_n}=n\,\mu/(\mu-1)$.
Some rescaling gives the large $n$ behaviour of the distribution
\begin{equation} 
  \label{eq:PiMuLarge}
   \pi_n(\xi)
   \simeq
   \frac{1}{(c_\mu n)^{1/\mu}}
   \,
   \mathcal{L}_{\mu,1}\left( \frac{\xi-\overline{\xi_n}}{(c_\mu n)^{1/\mu}} \right)
   \:,
\end{equation}
where $c_\mu=\Gamma(1-\mu)\,\cos(\pi\mu/2)>0$ and the L\'evy distribution $\mathcal{L}_{\mu,1}(x)$ was defined above, Eq.~\eqref{eq:LevyDistribution}. 
For some special values of $\mu$, explicit expressions are known.  
For instance one has 
for $\mu=3/2$
\begin{align}
  \mathcal{L}_{\frac32,1}(x) &= 
  \frac{\sqrt{3}}{\sqrt{\pi} |x|}
  \,
   e^{\frac{x^3}{27}}
   \times
   \begin{cases} 
   W_{\frac{1}{2}, \frac{1}{6}}
   \left(-\frac{2 x^3}{27}\right) 
   &\mbox{for } x \leq 0 
   \\[0.25cm]
   \frac{1}{6} W_{-\frac{1}{2},\frac{1}{6}}
   \left(\frac{2 x^3}{27}\right) 
    &\mbox{for }x \geq 0
   \end{cases} 
   \\   &
    \simeq
 \begin{cases} 
   \frac13\sqrt{\frac{-2 x}{\pi}} \,e^{\frac{2x^3}{27}}  & \mbox{for } x \to-\infty 
   \\[0.25cm]
   \frac{3}{2\sqrt{2\pi}\,x^{5/2}}                    & \mbox{for } x \to+\infty 
 \end{cases}
\end{align}
where $W_{\mu,\nu}(z)$ is the Whittaker function.
As in the case $\mu\in]0,1[$, the distribution presents a heavy tail only for positive $x$, however for $\mu\in]1,2[$ the support is not restricted to~$\mathbb{R}^+$ and there is a stretched exponential tail on the left side.

We emphasize that $\pi_n(\xi) \simeq \mu\,n \,\xi^{-1-\mu}$ for $\xi \to \infty$ holds for both $0 < \mu < 1$ and $1 < \mu < 2$.

This immediately gives the distribution of $G^\free(x)$~: 
$P_x^\free(G) = \pi_x(G)$.
In the fixed case, the arguments developed for $0 < \mu < 1$ also holds for $1<\mu<2$, leading to the far tail $P_x^\fixed(G)\sim x L \,G^{-1-2\mu}$.

\subsection{Marginal case \texorpdfstring{$\mu=1$}{TEXT}}

The case $\mu=1$ car also be studied along the same lines.
Eq.~\eqref{eq:LaplacePareto} now involves the exponential integral $\Gamma(0,s)=E_1(s)=\int_s^\infty(\D t/t)\,e^{-t}$, thus 
\begin{equation}
    \tilde q(s)
    \underset{s\to0}{\simeq }
    1 + s \,\left[\ln s + \mathbf{C}-1\right] + \mathcal{O}(s^2)
    \:,
\end{equation}
where $\mathbf{C}\simeq0.577$ is the Euler-Mascheroni constant. 
The substitution $s\to\I k$, required in order to deal with the Fourier transform, leads to 
\begin{align}   
\label{LaplacePimu1} 
  \tilde\pi_n(\I k)
  \simeq
  \exp\left\{
     -\I k n (1-\mathbf{C}) - n\frac{\pi}{2}|k|
     \left[
        1- \frac{2\I}{\pi}\,\,\mathrm{sgn}(k) \ln|k|
     \right]
  \right\}
\end{align}
for $k\to0$ and large $n$ with $k\,n\sim\mathcal{O}(1)$.
We recognize the Fourier transform of the L\'evy distribution of index $\mu=1$. In this case Eq.~\eqref{eq:LevyDistribution} is replaced by (see \cite{App04})~:
\begin{equation}
  \mathcal{L}_{1,\beta}(x)
  = \int_{\mathbb{R}} \frac{\D k}{2 \pi} e^{\I k x} 
  e^{ 
    -|k| \big[ 1-  \frac{2\I\beta}{\pi}\mathrm{sgn}(k)\ln|k| \big]
    }
\end{equation}
As a result, in the large $n$ limit, the distribution takes the form
\begin{equation}
    \pi_n(\xi) 
    \simeq
    \frac{2}{n\pi}\:
    \mathcal{L}_{1,1}
    \left(
      2\frac{\xi - \overline{\xi_n}}{n\pi}
    \right)
    \hspace{0.5cm}\mbox{where }
    \overline{\xi_n} = n\ln\left(n\frac{\pi}{2}e^{1-\mathbf{C}}\right)
    \:.
\end{equation}
This provides the distribution $P_x^\free(G) = \pi_x(G)$ in the marginal case $\mu=1$.
Hence $G^\free(x)\sim x\ln x$ in this case.


\section{Continuous model of interface}

The present model can be extended to a continuum setting. For instance 
consider the following model for an interface of height $h(x,t)$, 
$x \in [0,L]$ where the linear operator describes an inhomogeneous
medium with local diffusion coefficient $k(x)$ (which replaces the discrete spring constants $k_i$) and random biases $F(x)$
\begin{align}
    \partial_t h(x,t) = \partial_x (k(x) \partial_x  - F(x) ) h(x,t) + \eta(x,t) 
\end{align}
Then for the "fixed" case (Dirichlet boundary conditions) $h(0,t)=h(L,t)=0$
the equilibrium interface has the correlator
\begin{align}
&\langle h(x) h(x') \rangle = \frac{ \left( \int_0^x \frac{\D y}{k(y)} e^{ \int_y^x \D z \frac{F(z)}{k(z)}} \right) 
\left( \int_{x'}^L \frac{\D y}{k(y)} e^{ \int_y^{x'} \D z\frac{F(z)}{k(z)}} \right)  
}
{ \int_0^L \frac{\D y}{k(y)} e^{ \int_y^{x'} \D z\frac{F(z)}{k(z)}} } 
\nonumber\\
&\hspace{6cm}\mbox{for } x \leq x'
\end{align}
This is obtained by extending the present method, or alternatively 
using the standard representation of the Green's function at fixed energy of the inhomogeneous
diffusion operator in one dimension, taken at zero energy (see e.g. \cite{bouchaud1990classical})

One can also
define a "free" boundary condition by setting $k(x)=0$ for $x \geq L$ 
with
\begin{equation}
\langle h(x) h(x') \rangle 
=  \int_0^{\mathrm{min}(x,x')} \frac{\D y}{k(y)}\, e^{ \int_y^x \D z \frac{F(z)}{k(z)}} 
\end{equation}
The study of this class of models is deferred to future work.

\end{appendix}



\bibliographystyle{phreport}
\bibliography{biblio}




\end{document}